\documentclass[12pt,a4paper]{article}
\usepackage{amssymb}

\usepackage{graphicx}
\usepackage{amsmath}
\usepackage{a4}

\input{tcilatex}

\begin{document}

\title{The Goldberger- Treiman Discrepancy}
\author{N. F. Nasrallah\thanks{%
e-mail: nsrallh@cyberia.net.lb} \\
%EndAName
Faculty of Science, Lebanese University \\
Tripoli, Lebanon}
\date{}
\maketitle

\begin{abstract}
The Golberger- Treiman discrepancy $\Delta _{GT}=1-\frac{m_{N}g_{A}}{f_{\pi
}G_{\pi N}}$ is related to the asymptotic behaviour of the pionic form
factor of the nucleon obtained from baryonic \textrm{QCD} sum rules. The
result is $.015\lesssim \Delta _{GT}\lesssim .022.$
\end{abstract}

\bigskip

\bigskip

The Goldberger-Treiman relation (GTR) \cite{1} 
\begin{equation}
m_{N}g_{A}=f_{\pi }G_{\pi N}
\end{equation}
which relates the nucleon mass $m_{N}$, the axial-vector coupling constant
in $\beta -\mathrm{decay}$ $g_{A}$, the $\pi $decay constant $f_{\pi }$ and
the $\pi -N$ coupling constant $G_{\pi N}$ is one of the most remarkable
relations of hadronic physics. Explicit chiral symmetry breaking by the
quark masses leads to small corrections to the GTR, the Goldberger -Treiman
discrepancy (GTD) \cite{2} 
\begin{equation}
\Delta _{GT}=1-\frac{m_{N}g_{A}}{f_{\pi }G_{\pi N}}
\end{equation}
which arises from the coupling of the divergence of the axial vector current
to the $J^{p}=0^{-}$ continuum. The evaluation of $\Delta _{GT}$ has been
addressed recently in the framework of baryon chiral perturbation theory 
\cite{3}. On the experimental side $g_{A}=1.267\pm .004$ and $f_{\pi
}=92.42\,\mathrm{MeV}$ are known to enough precision and most of the
uncertainty in $\Delta _{GT}$ results from the uncertainty in $G_{\pi N}$.
The most recent determination of $G_{\pi N}$ from $NN,$ $N\bar{N}$ and $\pi
N $ data is by the Nijmegen group \cite{4} 
\begin{equation}
G_{\pi N}=13.05\pm .08\text{ which corresponds to }\Delta _{GT}=.014\pm .009
\label{3}
\end{equation}

Similar results are obtained by the VPI group \cite{5}. Larger values are
given by Bugg and Machleidt \cite{6} and by Loiseau et al. \cite{7} 
\begin{equation}
G_{\pi N}=13.65\pm .30\text{ which corresponds to }\Delta _{GT}=.056\pm .02
\end{equation}

The result of theoretical calculations at the loop level \cite{3} do not
account even for the smaller value given by eq. (\ref{3}) in a \ parameter
free way.

The evaluation of the GTD involves the integral over the imaginary part of
the form factor $\Pi (q^{2})$ which describes the matrix element of the
divergence of the axial current between two nucleon states 
\begin{eqnarray}
\langle P(p^{\prime }) &\mid &\partial _{n}A_{n}^{+}\mid N(p)\rangle =\Pi
(q^{2}).\mathcal{\bar{U}}(p^{\prime })\gamma _{5}\mathcal{U}(p) \\
q &=&p^{\prime }-p  \notag
\end{eqnarray}

Access to $\Pi (q^{2})$ is provided by the study of the three-point function 
\cite{8} 
\begin{equation}
\Gamma (t,q^{2})=-\iint d^{4}x\,d^{4}y\,\exp (-ipx)\exp (iqy)\,\langle 0\mid
T\Psi _{\sigma }^{P}(x)\partial _{\mu }A_{\mu }^{+}(y)\bar{\Psi}_{\kappa
}^{N}(0)\mid 0\rangle  \label{6}
\end{equation}
where $t=p^{2}$, $\partial _{\mu }A_{\mu }^{+}=i(m_{u}+m_{d})(\bar{u}\gamma
_{5}d)$ expresses the divergence of the axial currents in terms of quark
fields and 
\begin{eqnarray}
\Psi _{\sigma }^{P} &=&\epsilon _{ijk}u_{i}^{T}C\gamma _{\alpha }u_{j}\gamma
_{5}\gamma _{\alpha }d_{k}  \notag \\
\Psi _{\kappa }^{N} &=&\epsilon _{ijk}d_{i}^{T}C\gamma _{\alpha }d_{j}\gamma
_{5}\gamma _{\alpha }u_{k}
\end{eqnarray}
are the nucleon currents \cite{9}

The amplitude (\ref{6}) contains nucleon double and single pole
contributions as well as a non singular contribution of the continuum 
\begin{equation}
\Gamma (t,q^{2})=(\gamma _{5}\rlap/q)\left[ -\frac{\lambda _{N}^{2}\Pi
(q^{2})m_{N}}{(t-m_{N}^{2})^{2}}+\frac{c}{(t-m_{N}^{2})}+\cdots \right]
\label{8}
\end{equation}
where $c$ is the unknown coefficient of the single pole contribution and $%
\lambda _{N}$ represents the coupling of the nucleon to its current 
\begin{equation}
\langle 0\mid \Psi _{\sigma }^{P}\mid P\rangle =\lambda _{N}\mathcal{U}%
_{\sigma }
\end{equation}
and where we have limited ourselves to the tensor structure $\gamma _{5}%
\rlap/q$ for simplicity. Any other choice is of course a priori valid
provided it leads to stability of the calculation as will be shown to be the
case here.

The next step is to evaluate $\Gamma (t,q^{2})$ in QCD. To this end use is
made of the operator product expansion of the currents entering in eq.(\ref
{6}). The lowest dimensional operators, which provide the dominant
contributions at short distances are the unit operator and the operators $%
\bar{q}q$ and $G_{\mu \nu }G^{\mu \nu }(=GG)$. As we shall only use the
coefficient of $\gamma _{5}\rlap/q$ in the expansion of the currents
entering in (\ref{6}), the even dimensional operators $1$ and $GG$ will he
multiplied by the small quark mass $m_{q}$ and their contribution will be
greatly reduced as compared to the one of the odd dimensional operator $\bar{%
q}q.$ The contribution of the latter has been evaluated in the third of ref. 
\cite{8} 
\begin{eqnarray}
&&\Gamma ^{\mathrm{QCD}}(t,q^{2})  \label{10} \\
&=&\frac{-2m_{q}\langle \bar{q}q\rangle }{2\pi ^{2}}\left[ \frac{t}{q^{2}}%
\ln (-t)+\frac{1}{8}\ln (-t)-\frac{1}{4}tI_{0}(t,q^{2})+\frac{1}{4}\ln
(-q^{2})+\cdots \right] (\gamma _{5}\rlap/q)  \notag
\end{eqnarray}
with 
\begin{equation}
I_{0}(t,q^{2})=\int_{0}^{1}\frac{dx}{t-x(1-x)q^{2}}\ln \left( \frac{%
-x(1-x)q^{2}}{-t}\right)
\end{equation}
and $-2m_{q}\langle \bar{q}q\rangle $ given by the Gell-Mann-Oakes-Renner
relation 
\begin{equation}
-2m_{q}\langle \bar{q}q\rangle =f_{\pi }^{2}m_{\pi
}^{2}\;\;\;\;\;\;\;\;,\;\;\;\;\;\;\;\;\;f_{\pi }=.0924\mathrm{GeV}
\label{12}
\end{equation}

Expression (\ref{10}) holds for both $t$ and $q^{2}$ in the deep euclidean
region. The next step is to extrapolate to the nucleon mass-shell, i.e. to
obtain $\Pi ^{\mathrm{QCD}}(q^{2})$ from expressions (\ref{8}), (\ref{10})
and the analytic properties of $\Gamma (t,q^{2})$. This extrapolation to the
mass shell is done over a large interval of the variables $t$ and $q^{2}$.
The method of QCD sum rules provides a tool for such an extrapolation where
the approximations are well defined and where numerical stability of the
result provides a useful check of their validity. Note that the smallness of 
$\Delta _{GT}$ results from the smallness of the quark masses and is not
calculated as the difference of large numbers. If this were the case the
method of QCD sum rules would not be accurate enough to be reliable.

For $q^{2}$ fixed at a large negative value $\Gamma (t,q^{2})$ has a cut on
the positive $t$-axis starting at $t=(m_{N}+m_{\pi })^{2}$ in addition to
the nucleon pole structure exhibited in eq. (\ref{8}). Consider now the
Laplace type integral \cite{10} $\frac{1}{2\pi i}\int_{c}dt\,\exp \left( 
\frac{-t}{M^{2}}\right) \Gamma (t,q^{2})$ in the complex $t$-plane over a
closed contour $c$ consisting of a circle of large radius and two straight
lines above and below the cut which run from threshold to $R$, $M^{2}$ is
the usual 'Borel mass' parameter. The exponential provides convenient
damping of the contribution of the integral over the continuum. This
contribution is of course unknown and provides the main uncertainty in the
QCD\ sum rule approach. It could be greatly damped by decreasing the
parameter $M^{2}$, but, as is well known this enhances the contribution of
the higher order unknown terms \ in the operator product expansion. We hope
to obtain an intermediate range of $M^{2}$ for which the contribution of the
continuum and that of the higher order terms are both negligible. If these
approximations are adequate this will show up in the stability of the
result. We shall find out that this is the case. On the circle $\Gamma $ is
well approximated by $\Gamma ^{\mathrm{QCD}}$, except possibly for a small
region near the real axis.

We obtain then 
\begin{eqnarray}
&&\Pi ^{\mathrm{QCD}}(q^{2})+c^{\prime }(q^{2})M^{2}=\frac{M^{6}f_{\pi
}^{2}m_{\pi }^{2}}{2\pi ^{2}\lambda _{N}^{2}\exp \left( \frac{-m_{N}^{2}}{%
M^{2}}\right) m_{N}}\cdot  \notag \\
&&\left[ E_{1}\left( \frac{R}{M^{2}}\right) \frac{1}{q^{2}}+\frac{3}{8M^{2}}%
E_{0}\left( \frac{R}{M^{2}}\right) -\frac{q^{2}}{4M^{4}}\int_{0}^{1}dx%
\int_{0}^{R}\frac{dt\,\exp \left( \frac{-t}{M^{2}}\right) }{q^{2}-\frac{t}{%
x(1-x)}}\right]  \label{13}
\end{eqnarray}

with 
\begin{equation}
E_{i}\left( \frac{R}{M^{2}}\right) =\int_{0}^{\frac{R}{M^{2}}}x^{i}\exp
(-x)\,dx
\end{equation}

$\Pi (s=q^{2})$ is an analytic function in the complex $s$-plane except for
a simple pole at $s=m_{\pi }^{2}$ and a right hand cut running along the
positive real axis from $s=9m_{\pi }^{2}$ to $\infty $. 
\begin{equation}
\Pi (s)=\frac{-2f_{\pi }m_{\pi }^{2}G_{\pi N}}{s-m_{\pi }^{2}}+\cdots
\end{equation}
furthermore 
\begin{equation}
\Pi (s=0)=2m_{N}g_{A}
\end{equation}

Consider next the integral $\frac{1}{2\pi i}\int_{c^{\prime }}\frac{ds}{s}%
(s-m^{\prime 2})\Pi (s)$ where $m^{\prime }$ is a mass parameter and $%
c^{\prime }$ is a closed contour consisting of a circle of large radius $%
R^{\prime }$and two straight lines above and below the cut which run from
threshold to $R^{\prime }$\footnote{$R^{\prime }$ need of course not be
equal to $R$ but they are of the same order and any resonable difference
between them results only in negligible numerical effects so we take $%
R^{\prime }=R$ to simplify the notation.}. Cauchy's theorem implies 
\begin{eqnarray}
&&-m^{\prime 2}\cdot 2m_{N}\,g_{A}-2f_{\pi }G_{\pi N}(m_{\pi }^{2}-m^{\prime
2})  \notag \\
&=&\frac{1}{\pi }\int_{9m_{\pi }^{2}}^{R}\frac{ds}{s}(s-m^{\prime 2})\func{Im%
}\Pi (s)+\frac{1}{2\pi i}\oint \frac{ds}{s}(s-m^{\prime 2})\Pi ^{\mathrm{QCD}%
}(s)  \label{17}
\end{eqnarray}
where we have used $\Pi (s)=\Pi ^{\mathrm{QCD}}(s)$ on the circle.

The first term on the r.h.s. of eq.(\ref{17}) represents an integral over
the unknown continuum. As $m^{\prime 2}$ is varied between threshold and $R$
this integral changes sign which implies that it vanishes for some value of $%
m^{\prime 2}$ which we adopt. Because $m^{\prime }$ is an unknown parameter
that we shall vary within reasonable limits it is superfluous to include any
contribution of the continuum near threshold. The GTD then follows from eq.(%
\ref{17}) 
\begin{equation}
\Delta _{GT}=\frac{m_{\pi }^{2}}{m^{\prime 2}}+\frac{1}{2f_{\pi }G_{\pi
N}\,m^{\prime 2}}\cdot \frac{1}{2\pi i}\oint \frac{ds}{s}(s-m^{\prime 2})\Pi
^{\mathrm{QCD}}(s)
\end{equation}

And when expression (\ref{13}) for $\Pi ^{\mathrm{QCD}}$ is used 
\begin{eqnarray}
\Delta _{GT} &=&\frac{m_{\pi }^{2}}{m^{\prime 2}}\bigg[1+\frac{1}{4\pi
^{2}G_{\pi N}}\left( \frac{f_{\pi }}{m_{N}}\right) \bigg(E_{1}\left( \frac{R%
}{M^{2}}\right) -\frac{3}{8}\frac{m^{\prime 2}}{M^{2}}E_{0}\left( \frac{R}{%
M^{2}}\right)  \label{19} \\
&&-\frac{1}{4}\int_{0}^{1}dx\int_{0}^{x(1-x)\frac{R}{M^{2}}}dy\exp
(-y)\left( \frac{y}{x(1-x)}-\frac{m^{\prime 2}}{M^{2}}\right) \bigg)\bigg]%
+c^{\prime \prime }M^{2}  \notag
\end{eqnarray}

$\lambda _{N}^{2}$ is obtained in a similar fashion from a study of the
nucleonic two point function $\int d^{4}x\exp (iqx)\langle 0\mid T\,\Psi
(x)\Psi (0)\mid 0\rangle $ \cite{9} with the result 
\begin{equation}
(2\pi )^{4}\lambda _{N}^{2}\exp \left( \frac{-m_{N}^{2}}{M^{2}}\right) =%
\frac{M^{6}}{4}E_{2}\left( \frac{R}{M^{2}}\right) -\frac{\pi ^{2}}{2}\langle 
\frac{\alpha _{s}GG}{\pi }\rangle M^{2}E_{0}\left( \frac{R}{M^{2}}\right) +%
\frac{32}{3}\pi ^{4}\langle (\bar{q}q)^{2}\rangle
\end{equation}

The choice of $M^{2}$ in eq.(\ref{19}) as well as the consistency of the
method is dictated by stability considerations. If there are values of $%
M^{2} $ small enough to provide adequate damping of the continuum and large
enough to justify the neglect of the contributions of higher order
condensates in the operator product expansion this should show up in the
stability of expression (\ref{19}). This means that the first term on the
r.h.s. of (\ref{19}) should show a linear behaviour which compensates the
linear variation of $c^{\prime \prime }M^{2}$ in some intermediate range of $%
M^{2}$ (Note that the curve need show no horizontal plateau, this happens
only if $c^{\prime \prime }=0$). The value of $m^{\prime 2}$ is expected to
be close to (albeit smaller because of the weight \ factor $\frac{1}{s}$)
the maximum of the $\pi ^{\prime }(1.7\mathrm{GeV}^{2})$ bump. It seems
reasonable to vary it in the range $1\mathrm{GeV}^{2}\lesssim m^{\prime
2}\lesssim 1.5\mathrm{GeV}^{2}$. For the gluon condensate we use the
standard value $\langle \frac{\alpha _{s}GG}{\pi }\rangle =.012\,\mathrm{GeV}%
^{2}$. For $\langle (\bar{q}q)^{2}\rangle $ the choice $\langle \bar{q}%
q\rangle ^{2}$ (vacuum saturation hypothesis) is usually made but as this
seems to be too stringent an assumption \cite{10}, we take $\langle
qq^{2}\rangle =\beta \langle qq\rangle ^{2}.$Varying $\beta $ between $1$
and $3$ has no noticeable effect on the result.

In the figure the first term on the r. h. s. of eq.(\ref{19}) is plotted
against $M^{2}$ for $m^{\prime 2}=1\mathrm{GeV}^{2}$ and $\beta =1$. It
clearly exhibits a slow linear variation in the range $.5\mathrm{GeV}%
^{2}<M^{2}<1.5\mathrm{GeV}^{2}$ which gives 
\begin{equation*}
\Delta _{GT}=.022
\end{equation*}
varying $m^{\prime 2}$as discussed above yields finally 
\begin{equation}
.015\lesssim \Delta _{GT}\lesssim .022
\end{equation}
which is consistent with the value given by eq.(\ref{3}) and clearly favours
the smaller value of $G_{\pi N}$.

It is finally worth investigating the possibility that the value of the
quark condensate $\langle \bar{q}q\rangle $ is much smaller than what
results from the GOR relation eq.(\ref{12}). This is the case for example in
''generalized Chiral Perturbation theory'' \cite{11}. We would then have 
\begin{equation}
\Delta _{GT}\simeq \frac{m_{\pi }^{2}}{m^{\prime }}\;\;\;\;\;\;\;\;\;\;\;\;\;%
\text{or \ \ \ \ \ \ \ \ \ \ \ \ \ \ }.10\lesssim \Delta _{GT}\lesssim .14
\end{equation}
\pagebreak

\end{document}